\documentclass[final,3p,times,twocolumn,nonatbib]{elsarticle}
\usepackage[utf8]{inputenc}
\usepackage[british]{babel}
\usepackage{csquotes}
\usepackage[backend=biber,style=apa]{biblatex}
\DeclareLanguageMapping{british}{british-apa}
\usepackage[T1]{fontenc}
\usepackage[utf8]{inputenc}
\usepackage{textcomp}
\usepackage{newtxtext,newtxmath}
\usepackage{graphicx,booktabs,array,tabularx,longtable,rotating}
\usepackage{microtype}
\usepackage{xurl}
\usepackage[dvipsnames]{xcolor}
\usepackage[colorlinks=true,linkcolor=MidnightBlue,citecolor=MidnightBlue,urlcolor=MidnightBlue,breaklinks=true]{hyperref}
\usepackage{caption}
\DeclareUnicodeCharacter{2212}{\ensuremath{-}}
\DeclareUnicodeCharacter{03C1}{\ensuremath{\rho}}
\DeclareUnicodeCharacter{0394}{\ensuremath{\Delta}}
\DeclareUnicodeCharacter{03B2}{\ensuremath{\beta}}
\DeclareUnicodeCharacter{00B2}{\textsuperscript{2}}
\DeclareUnicodeCharacter{00D7}{\ensuremath{\times}}
\DeclareUnicodeCharacter{2248}{\ensuremath{\approx}}
\DeclareUnicodeCharacter{2264}{\ensuremath{\leq}}
\DeclareUnicodeCharacter{2265}{\ensuremath{\geq}}
\DeclareUnicodeCharacter{00B0}{\textdegree}
\DeclareUnicodeCharacter{2010}{-}
\DeclareUnicodeCharacter{2076}{\textsuperscript{6}}
\DeclareUnicodeCharacter{00B1}{\ensuremath{\pm}}
\DeclareUnicodeCharacter{2020}{\textdagger}
\newcommand{\docxellipsis}{.\kern\fontdimen3\font.\kern\fontdimen3\font.}
\newcolumntype{L}[1]{>{\raggedright\arraybackslash}p{#1}}
\newcolumntype{Y}{>{\raggedright\arraybackslash}X}
\usepackage{flushend}
\journal{Preprint}
\makeatletter
\def\ps@pprintTitle{\let\@oddhead\@empty\let\@evenhead\@empty\def\@oddfoot{\footnotesize\itshape Preprint\hfill 8 November 2026}\let\@evenfoot\@oddfoot}
\makeatother
\hypersetup{pdftitle={Trail Making Tests in Extended Reality: A Critical Review with Implications for Cognitive Ageing},pdfauthor={Panagiotis Kourtesis},pdfsubject={Preprint},pdfkeywords={Trail Making Test, Color Trails Test, extended reality, virtual reality, neuropsychological assessment, cognitive training, cognitive ageing, older adults, mild cognitive impairment, interaction techniques}}
\begin{document}
\begin{frontmatter}
\title{Trail Making Tests in Extended Reality: A Critical Review with Implications for Cognitive Ageing}
\author[aff1,aff2,aff3]{\href{https://orcid.org/0000-0002-2914-1064}{Panagiotis Kourtesis}\corref{cor1}}
\address[aff1]{Department of Psychology, The American College of Greece, 153 42 Athens, Greece; pkourtesis@acg.edu}
\address[aff2]{Department of Psychology and Department of Informatics and Telecommunications, National and Kapodistrian University of Athens, 157 72 Athens, Greece}
\address[aff3]{Department of Psychology, University of Edinburgh, Edinburgh EH8 9JZ, UK}
\cortext[cor1]{Correspondence: \href{mailto:pkourtesis@acg.edu}{pkourtesis@acg.edu}}
\begin{abstract}
Extended reality (XR) versions of the Trail Making Test and Color Trails Test are promising for assessing and training cognition in older adults. However, each XR adaptation is a new instrument, whose display, interaction technique, timing rules and layouts may change what it measures. This critical narrative review maps these implementations by software lineage and appraises their evidence for assessment and training against common criteria. The Trail Making Test in Virtual Reality (TMT-VR), whose source code was inspected, serves as a measurement case. XR trail-making tests showed convergence with paper and computerised versions, varying by test part and sample, and clinical group differences (e.g., Parkinson's disease, mild cognitive impairment). However, gaps in reliability (e.g., randomised forms), validation (e.g., norms, prospective samples) and reporting (e.g., error handling, build identity) emerged, and no training benefit attributable to the trail-making component was identified. The TMT-VR case shows how interaction technique, timing rule, error handling and unlogged layouts shape a score's meaning. Future tests should log layouts and events, estimate alternate-form reliability and undergo prospective validation, and training trials should isolate the trail-making component. Several XR trail-making tests have moved beyond feasibility, and reported tolerability during supervised sessions supports further evaluation in older adults.
\end{abstract}
\end{frontmatter}

\section{Introduction}\label{sec:1}

Cognitive assessment and cognitive training in older adults both need tasks that are interpretable, acceptable and accessible. Immersive virtual reality (VR) is attractive for both purposes, because it adds experimental control, automated scoring and everyday-like content to a task \parencite{ref1,ref2}. However, every adaptation of a paper test is a new instrument, which needs its own reliability, validity, norms and technical documentation before a clinical claim is made \parencite{ref3}. It also needs software of adequate quality: suitable hardware, ergonomic interaction, tutorials, a steady frame rate and a measured symptom outcome \parencite{ref4,ref5,ref6}. The field has not yet met these standards in general. Across 287 studies of VR neuropsychological testing, three tools had normative data \parencite{ref7}, and across 24 studies of VR memory assessment, only one made its tasks openly accessible \parencite{ref8}.

Older adults are not one group. Healthy volunteers in their sixties, adults over 80, people with mild cognitive impairment (MCI) or dementia and care-home residents differ in vision, hearing, motor control, posture, fatigue and digital experience \parencite{ref9,ref10}. Each of these characteristics can change a score or the willingness to continue. Head-mounted VR has generally been accepted by older adults, symptoms have usually been minor, and empirically grounded design guidance exists \parencite{ref11,ref12,ref13,ref14,ref15}. Nonetheless, feasibility in young adults is not evidence for clinical ageing, and acceptability is not evidence of validity. Training raises a further distinction. Assessment validity shows that a task measures a construct; it does not show that practising the task improves cognition. The evidence on VR training for MCI also comes from heterogeneous intervention packages, in which the contribution of any single task is hard to isolate \parencite{ref16,ref17}.

Trail making is a revealing case for these questions. The Trail Making Test (TMT) began as a subtest of the Army Individual Test Battery \parencite{ref18} and entered clinical use as an indicator of brain damage \parencite{ref19}. Part A requires the examinee to connect 25 numbered circles in order, and Part B requires alternation between numbers and letters (i.e., 1-A-2-B \docxellipsis{} 13). The examiner points out each error as it occurs, and the clock keeps running \parencite{ref20}. Part A time is carried mainly by perceptual speed and visual search, with a smaller motor share \parencite{ref21,ref22,ref23}. Part B adds working memory, set switching and symbol handling \parencite{ref24,ref25}, as well as a longer and more cluttered sheet \parencite{ref26}. Age is the dominant demographic factor, education matters for Part B in later life, and errors add sensitivity to MCI \parencite{ref27,ref28,ref29}. The Color Trails Test (CTT) replaces letters with colour alternation to reduce the dependence on language and literacy \parencite{ref30,ref31}. Computerised, tablet, walking, driving and screen-based eye-tracking versions already exist and set the psychometric bar for immersive versions (Supplementary Section S2) \parencite{ref32,ref33,ref34,ref35,ref36,ref37,ref38,ref39,ref40,ref41,ref42,ref43,ref44}.

Extended reality (XR) changes the trail-making task in five places (Figure 1). The display adds depth, visual angle and head movement \parencite{ref1}. The response becomes gaze, head orientation, a hand-held pointer or touch, each with its own speed and error profile \parencite{ref45,ref46}. The confirmation rule (i.e., a dwell, a trigger or a touch) decides how an incidental pass over a wrong target is treated. The geometry can change at every administration, which reduces the memorisation of a fixed form but adds a layout component to every score. Finally, XR can provide timestamped records of selections, gaze and head pose, although these records are not implemented or reported consistently. A paper examiner also records errors and times, so the potential gain is one of resolution, consistency and auditability \parencite{ref47}.

\begin{figure*}[!tp]
\centering
\includegraphics[width=0.82\textwidth]{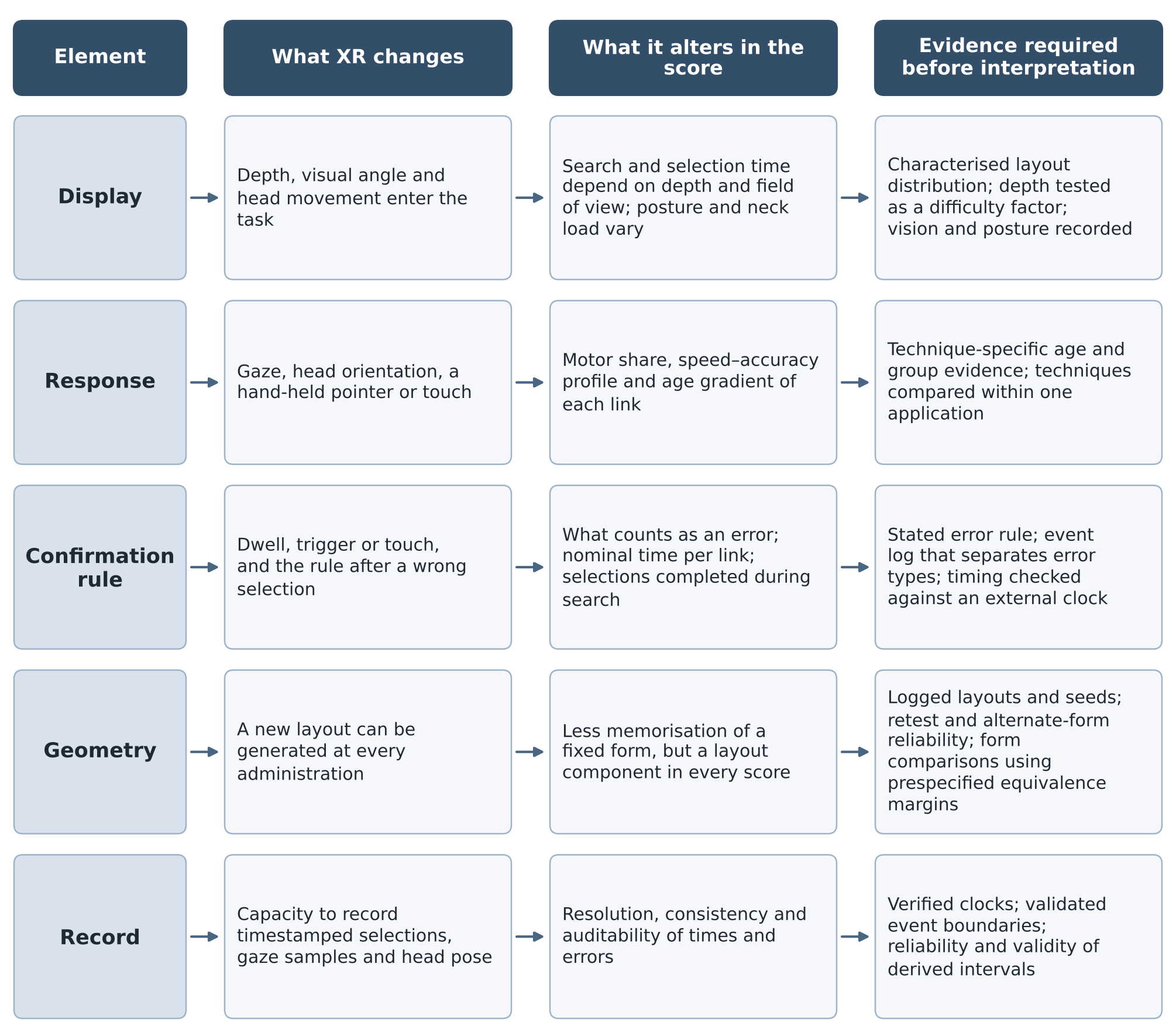}
\caption{What extended reality changes in a trail-making test and the evidence each change requires. Each row names a change, what it alters in the score and the evidence needed before the score can be interpreted.}\label{fig:1}
\end{figure*}

This review asks what the evidence on XR trail-making tests supports for cognitive assessment and training, particularly in older adults, and which design and measurement requirements remain unmet. Eligibility is all-age, because much of the relevant work involves younger adults, and the limits of transfer to older and clinical groups are stated where they apply. The review had three aims. The first was to map and appraise the implementations by software lineage, applying the same criteria to each. The second was to use the Trail Making Test in Virtual Reality (TMT-VR), whose source code and study records were available for inspection, as a measurement case. The third was to state what XR trail tests must demonstrate before clinical use or training claims, for readers in neuropsychology and in human–computer interaction (HCI). Statements that a property was not found refer to the sources identified by this review (Section 2).

\section{Review Methods}\label{sec:2}

This is a critical narrative review with a structured evidence map, assembled from the project library and targeted supplementary searching. It is not a systematic or scoping review. The synthesis is narrative, because the question concerns design and evidence quality rather than a pooled effect.

Records were eligible for the XR inventory if they described a trail-making task delivered in head-mounted VR, head-mounted augmented or mixed reality (AR/MR), or a dome or large-screen immersive display. A trail-making task was defined as connecting targets in a prescribed sequence defined by numbers, letters, colours or their alternation, including the CTT and its relatives. Assessment and training uses, any population, any publication type, any date and any language were eligible. Conventional, desktop, tablet, walking, driving and screen-based eye-tracking trail tests were retained as contextual comparators and were not counted in the inventory. A study that used a conventional TMT only as a comparator for immersive tasks of another kind was excluded \parencite{ref48}. Studies in which a conventional TMT served only as the outcome of an XR intervention were classed as training evidence, unless the intervention itself contained a trail-making task.

The library was assembled by the author between September and 24 October 2026 from database searches and from the reference lists of held papers, using trail-making, immersion and training terms. The query strings and per-database counts were not logged, so the search cannot be reproduced as a database protocol. Supplementary Section S1 reports it instead by its intakes, dates and record counts, with a reconciliation of the candidate records that a gap analysis identified. The evidence map holds 152 records with 37 fields (File S1). Of these, 144 were read in full text or in named sections, five only as abstracts or publisher records, two only by title and one only as a library note. Records known only from an abstract were included at that level and are marked as such in the tables. Screening and charting were performed by the author, and no second screener was involved. Generative AI tools assisted with organising the literature records and with extracting and verifying their data, as detailed in the Acknowledgments.

Four units were kept distinct, because conflating them inflates the apparent evidence: the implementation lineage (related software from one group), the configuration (a version with its hardware, input, timing and language), the publication and the study. Every implementation was appraised against the same criteria in five domains. Reporting completeness used the RATE-XR items as reporting guidance and not as validation \parencite{ref6}. The other domains were study method, measurement evidence \parencite{ref3}, technical verification of timing, scoring, builds and layouts \parencite{ref5,ref49}, and human factors and applicability to older adults and clinical groups \parencite{ref50,ref51}. Each criterion was recorded as reported, not reported, not assessed, not applicable or an observed shortcoming. Criteria were scaled to the use an implementation claims, so diagnostic norms were not demanded of a game evaluated only for acceptability. No compliance score was computed. The full matrices by criterion are Supplementary Tables S7.1 and S7.2, and the charting schema is Table S2.2. The author developed the TMT-VR. It was charted and appraised with the same schema and criteria as every other implementation.

\section{Trail Making in Extended Reality}\label{sec:3}

\subsection{Implementation Lineages and Human Factors}\label{sec:3.1}

Table 1 groups the implementations by lineage. Two assessment lineages have been evaluated across several studies. The VR Color Trails Test (VR-CTT) of the Sheba group has eight reports read in full and a ninth known only from its abstract. It translates the paper CTT into a fixed three-dimensional layout that the participant touches with a hand-held pointer while standing, first in a 360° dome and then in a head-mounted display \parencite{ref52,ref53,ref54}. The three-dimensional trajectory of the pointer has been recorded since the first report \parencite{ref52}. Later reports added analyses of head–hand coordination, gaze and pupil size \parencite{ref47,ref55}, an eyes-only mode in which a fixation selects the target \parencite{ref47}, shortened forms \parencite{ref56,ref57} and a study in Parkinson's disease \parencite{ref58}. The ninth report used the VR-CTT, with an added mirror-reversal module, as an outcome measure in 13 medical residents assessed around a 26-h night shift \parencite{ref59}. The second lineage is the TMT-VR, with nine reports on three configurations (Section 4). A motion-capture assessment system includes shortened CTT Parts A and B among its tasks and has one report \parencite{ref60}. The AR-TMT places targets on physical panels in a 180° arc around a seated participant, under spacing constraints and distraction conditions \parencite{ref61,ref62}. A VR-TMT for epilepsy screening keeps all targets at 1 m on a curved surface \parencite{ref63}. The remaining entries are an immersive TMT delivered in one system with VR cognitive rehabilitation \parencite{ref64} and a VR TMT-B game demonstrated without participants \parencite{ref65}. An AR music-making and painting application includes a trail-making mode to assess attention \parencite{ref66}, and the GAME2AWE exergame platform includes the Darts game \parencite{ref67,ref68}. Two further implementations are known only from a technical report and an abstract \parencite{ref69,ref70}. Of the two games, the VR TMT-B game pairs letters with animals and spreads its targets over 60–210°, and its layout algorithm varies target locations each time a level is loaded \parencite{ref65}. The GAME2AWE platform lists target randomisation among its gamification elements, without specifying it for the Darts game \parencite{ref67}.

\begin{table*}[!tp]
\caption{XR trail-making implementations by lineage (full inventory with page anchors in Supplementary Section S3). Reported study samples are shown; analysis denominators vary where data were incomplete. ADHD, attention-deficit/hyperactivity disorder; n/s, not stated.}\label{tab:1}
\centering\scriptsize\setlength{\tabcolsep}{3pt}\renewcommand{\arraystretch}{1.12}
\begin{tabular}{L{0.1250\dimexpr\linewidth-42pt\relax}L{0.1650\dimexpr\linewidth-42pt\relax}L{0.1500\dimexpr\linewidth-42pt\relax}L{0.1500\dimexpr\linewidth-42pt\relax}L{0.1650\dimexpr\linewidth-42pt\relax}L{0.1450\dimexpr\linewidth-42pt\relax}L{0.1000\dimexpr\linewidth-42pt\relax}}
\toprule
\textbf{Lineage or implementation} & \textbf{Reports} & \textbf{Display; posture} & \textbf{Selection; confirmation} & \textbf{Task; layout} & \textbf{Samples} & \textbf{Use} \\
\midrule
VR-CTT, dome & \cite{ref52,ref54} & CAREN 360° dome, tracked wand; standing & reach-touch & CTT Parts A and B; fixed layout in depth & 11 (pilot); 43 aged 18–64 & assessment \\
VR-CTT, head-mounted & \cite{ref47,ref53,ref54,ref55,ref56,ref57,ref58} & HTC Vive with controller; eye tracker in one study; standing & controller reach-touch; eyes-only fixation in one study & CTT Parts A and B; fixed layout in depth; shortened forms & 56; 104 aged 18–90 (2 did not complete); 122; 30 older adults; 23 with Parkinson's disease and 23 controls; 15 (forms); 165 across head-mounted, dome and tablet formats (retrospective short forms) & assessment \\
VR-CTT, abstract-level report & \cite{ref59} & n/s & three-dimensional manual movements & CTT with an intermittent mirror-reversal module; six sessions & 13 medical residents & outcome measure in an occupational study \\
Motion-capture CTT & \cite{ref60} & VIVE Pro Eye with motion-capture gloves; n/s & glove-tracked hand touch & shortened CTT A and B among other tasks & 50 aged 19–24 & assessment \\
AR-TMT & \cite{ref61,ref62} & Magic Leap 2, optical see-through; seated & controller & TMT A and B on physical panels; constrained random positions; distraction & 34 aged 18–44 & assessment \\
TMT-VR, Varjo Aero PC, 2024–2025 & \cite{ref71,ref72,ref73} & Varjo Aero; standing in one study, n/s in two & eye, head or controller ray-casting; dwell reported as 2 s or 1.5 s & TMT A and B; random layout per run, in depth & 71 aged 18–35; 53 aged 18–40 (25 with ADHD); 77 aged 19–56 & assessment \\
TMT-VR, VIVE Pro Eye PC, English, 2025 & \cite{ref74,ref75} & VIVE Pro Eye; standing & head-gaze; 0.5-s dwell & TMT A and B; random layout per run, in depth & 91 (47 aged 20–38; 44 aged 64–81) & assessment \\
TMT-VR, Focus Vision standalone, Greek, 2026 & \cite{ref76,ref77,ref78,ref79} & VIVE Focus Vision; seated & head-gaze; 0.5-s dwell & TMT A and B; random layout per run, 3–8 m & 202 aged 19–90; 230 aged 19–90; 84 aged 60–89, 42 with MCI & assessment \\
VR-TMT for epilepsy screening & \cite{ref63} & Meta Quest 3 with eye tracker; seated & controller button; wrong selections do not advance & TMT A and B on a curved surface at 1 m; no depth & 8 aged 23–58 & assessment \\
Immersive TMT in cancer rehabilitation & \cite{ref64} & n/s & n/s & TMT A and B within a rehabilitation system & 9 of 17 eligible completed & assessment and training outcome \\
VR TMT-B game & \cite{ref65} & Vive Pro Eye & pointing and click & letters and animals, 8–16 targets over 60–210°; locations vary each time a level is loaded & none & demonstration \\
AR trail-making mode (SoundShaping) & \cite{ref66} & HoloLens 2; n/s for the trail mode & pinch-draw; a wrong node does not advance & fixed, eight numbered nodes & not yet evaluated & embedded assessment of attention (planned) \\
Darts game, GAME2AWE & \cite{ref67,ref68} & Meta Quest 2; n/s & head-gaze crosshair held on the target & 12 balloons, numbered and lettered, in order & 8 and 15 older adults, with 4 experts (feasibility); 58 in a platform trial & training content \\
VR TMT (technical report) & \cite{ref69} & n/s & n/s & TMT A and B & n/s (abstract level) & assessment \\
VR-TMT on an Android headset & \cite{ref70} & n/s & n/s & TMT A and B & n/s (abstract level) & assessment \\
\bottomrule
\end{tabular}
\end{table*}

The implementations differ in every element that Figure 1 identifies. Posture ranges from standing, secured in the dome or free in a tracked area, to sitting on a fixed or swivel chair. Selection ranges from reaching and touching, through a controller ray and button, to gaze or head orientation held on the target for a dwell. Confirmation rules differ even within a lineage. The eyes-only VR-CTT, for example, required a 100-ms fixation to select a correct target and a 500-ms fixation to select a wrong one, so that wrong selections were harder to complete \parencite{ref47}. Layouts are either fixed, as in the paper test, or newly generated, and either spread in depth or held at one distance. The AR-TMT randomises target positions under spacing constraints and checked its arrangements by the mean head movement per target, which had a coefficient of variation of 7.7\% across stages \parencite{ref61}. Target fields also differ in angular extent. The VR-CTT spreads its targets over about 120° horizontally, wider than the headset's field of view, and 52–62° vertically, so that head movement is part of the search \parencite{ref47,ref55}. The TMT-VR spans about 80° horizontally (Supplementary Section S6), whereas the epilepsy VR-TMT keeps all targets at one distance in the forward field, so that they share visual angle and selection conditions \parencite{ref63}. Error handling also differs. In the VR-CTT and the TMT-VR, wrong selections are registered and counted, and the participant continues from the last correct target. In the epilepsy VR-TMT, the motion-capture CTT, the VR TMT-B game and the SoundShaping trail mode, a wrong selection does not advance the trail, and the participant must select the correct target \parencite{ref60,ref63,ref65,ref66}. The AR-TMT logs each selection as a hit or a miss without stating whether a miss advances the trail \parencite{ref61}. The reports of the Darts game and the cancer-care TMT do not describe error handling. A rule that stops a wrong selection from advancing the trail does not stop the selection itself, so completion times and error counts are not comparable across implementations unless each states whether wrong inputs are registered, counted, corrected or ignored.

These choices bear directly on older users. Design guidance for immersive VR with older adults emphasises onboarding and assistance, practice, safe seating and multimodal feedback \parencite{ref14,ref15}. In the empirically grounded set, first-use interviews gave no support to the advice to limit head movement or to avoid buttons and gestures, and some older adults found seated use limiting \parencite{ref14}. This guidance says nothing about gaze or dwell selection, albeit one review notes that age-related decline of the vestibulo-ocular reflex may complicate gaze-based interaction \parencite{ref10}. A design feature is therefore alignment with a recommendation, not evidence that the recommendation was met for intended users. Head-gaze selection, for example, follows the advice to avoid buttons, whereas its head movement departs from the advice to limit head movement, and it places a demand on neck mobility. Fitting is reported unevenly too. Among the TMT-VR studies, one reports automatic interpupillary-distance adjustment \parencite{ref76}, and another reports that the examiner fitted the headset and adjusted the interpupillary distance \parencite{ref79}. Depth deserves particular care. It costs selection time in three-dimensional pointing \parencite{ref46,ref80}, and an incorrectly set interpupillary distance can impair depth perception \parencite{ref10}. Older players also found depth difficult when reaching in one exergame, in a sample that excluded cognitive and visual impairment \parencite{ref81}. A test that uses depth therefore needs a characterised layout distribution and depth treated as a candidate difficulty factor.

\subsection{Measurement Properties and Clinical Evidence}\label{sec:3.2}

Convergent evidence against a paper or computerised trail test was reported, with checkable estimates, for four implementations, and its size depended on the test part, the comparator and the sample (Table 2). The VR-CTT correlated moderately to strongly with the paper CTT on both parts in healthy adults, and numerically less strongly in patients with Parkinson's disease, albeit without a significant difference between the groups' coefficients \parencite{ref54,ref58}. The first headset estimates came from 56 healthy young and middle-aged adults who stood and used a controller, with Pearson coefficients pooled across the two age groups \parencite{ref53}. The motion-capture CTT correlated more strongly on Part A than on Part B in young adults \parencite{ref60}. The AR-TMT correlated with a monitor-based TMT on Part B only, and an in-headset motor-skill score correlated with its Part A and not with the monitor Part A, which is consistent with a larger motor share in the AR format \parencite{ref61}. The eye-tracking TMT-VR correlated weakly with the paper test on both parts \parencite{ref72}. The head-gaze version correlated moderately on both parts in a sample of younger and older adults \parencite{ref75}, and its Part B correlation was numerically lower in older adults with MCI than in their comparison group \parencite{ref79}. A Japanese report described significant correlations whose sample and coefficients could not be checked \parencite{ref69}. These associations therefore vary by implementation, part and population, and no study has isolated why. The estimates also differ in kind. Some are Spearman and some Pearson coefficients, and correlations pooled across age groups can partly reflect the separation of the groups. Convergence shows that the formats share information. It does not establish agreement, interchangeable scores or clinical use \parencite{ref3}.

Retest reliability was estimated for one lineage, the VR-CTT, on a fixed layout \parencite{ref54,ref58}. Repeated administration also brings practice effects. In the residents' study, completion times improved across six sessions, mainly in the mirror-reversal segments, whereas regular segments stabilised after pre-shift practice \parencite{ref59}. The VR-CTT's shortened forms illustrate how far the reliability question remains open. In 15 young adults who completed five counterbalanced forms, the reported tests did not detect differences between forms or orders \parencite{ref57}. That result establishes neither equivalence nor alternate-form reliability. The first shortened versions were derived retrospectively from complete 25-target trials of an existing dataset that combined head-mounted, dome and tablet administrations \parencite{ref56}. Their correlations with the full test therefore share observations and cannot independently validate the short forms. The associations of these forms with the paper CTT and with age were also estimated in the same data from which they were derived. No retest or alternate-form estimate was identified for any randomised-layout XR trail test.

Clinical group evidence was reported for two lineages. The VR-CTT distinguished patients with Parkinson's disease from matched controls in completion time, errors and kinematic measures \parencite{ref58}. The TMT-VR distinguished adults with MCI from comparison adults on Part B scores, and no statistically significant difference in diagnostic accuracy from the paper test was detected after correction \parencite{ref79}. The eye-tracking TMT-VR found no completion-time difference between adults with attention-deficit/hyperactivity disorder (ADHD) and controls, but more wrong selections in ADHD \parencite{ref72}. Age-group discrimination was close to that of the paper test for the VR-CTT \parencite{ref54} and better than the paper test on Part B scores for the English TMT-VR \parencite{ref75}. Age-group discrimination is not evidence of clinical accuracy. No XR trail test was found to have normative data, a prospective validation in a referred clinical population or evidence of incremental value over an existing screening instrument.

Process measures are the most distinctive evidence of the VR-CTT lineage. An early headset report fitted a planar minimal-jerk model to selected movement paths, 14 in Part A and 12 in Part B \parencite{ref53}. In a later study of 30 older adults, search accounted for 58\% of completion time in Part A and 74\% in Part B \parencite{ref47}. These older adults differed from a small young group in search duration and not in execution duration or velocity. Head–hand lag lengthened from Part A to Part B and with age \parencite{ref55}, and patients with Parkinson's disease differed from controls in peak hand velocity and head–hand lag rather than in the share of search \parencite{ref58}. These results show what continuous hand, head and gaze records can add to completion time. They come from fixed layouts, and in the clinical study the discriminating variables were selected in the same sample without cross-validation. Error scores are less developed. Patients with Parkinson's disease made more errors than controls in both formats \parencite{ref58}, and TMT-VR errors were age-related and separated adults with MCI (Table 2). Elsewhere, errors were logged but not reported \parencite{ref61}, constrained by design \parencite{ref63} or not scorable in the eyes-only mode \parencite{ref47}.

Usability and tolerability were measured for the TMT-VR, the motion-capture CTT and the Darts game, with ad hoc items for the epilepsy VR-TMT and with stage ratings of comfort for the AR-TMT. The VR-CTT reports no questionnaire, and 2 of its 104 lifespan participants, both older adults, did not complete the head-mounted version \parencite{ref54}. Symptoms, where measured, were generally mild, in line with the older-adult literature \parencite{ref13}. Most were measured after headset removal, for example with the Cybersickness in Virtual Reality Questionnaire (CSQ-VR) \parencite{ref82}. These results are human-factors evidence. They show that a format was tolerated under supervision; they do not show that its scores measure the intended construct.

\begin{table*}[!tp]
\caption{Measurement evidence for the XR trail-making implementations. Estimates are given with their comparator, estimator and sample. AUC, area under the receiver operating characteristic curve; g, Hedges' g; ICC, intraclass correlation coefficient; M-ACE, Mini-Addenbrooke's Cognitive Examination; n/r, not reported in the sources identified; SUS, System Usability Scale.}\label{tab:2}
\centering\scriptsize\setlength{\tabcolsep}{3pt}\renewcommand{\arraystretch}{1.12}
\begin{tabular}{L{0.1400\dimexpr\linewidth-36pt\relax}L{0.1700\dimexpr\linewidth-36pt\relax}L{0.2200\dimexpr\linewidth-36pt\relax}L{0.1200\dimexpr\linewidth-36pt\relax}L{0.1900\dimexpr\linewidth-36pt\relax}L{0.1600\dimexpr\linewidth-36pt\relax}}
\toprule
\textbf{Implementation} & \textbf{Usability or tolerability} & \textbf{Convergent evidence (comparator; sample)} & \textbf{Reliability} & \textbf{Group differences; diagnostic accuracy} & \textbf{Norms; prospective validation} \\
\midrule
VR-CTT \parencite{ref52,ref53,ref54,ref58,ref59} & no questionnaire; 2 of 104 participants, both older adults, did not complete the head-mounted test & paper CTT: dome pilot ρ .76 (A), .88 (B), n = 11; dome .58/.70, n = 43 with one Part A value missing, and headset .62/.69, n ≈ 102, estimator labelled inconsistently; headset r .63/.60, n = 56; ρ .6/.7 controls and .4/.5 Parkinson's disease, n = 45 & ICC .60/.68 at 12 weeks (dome); .59–.89 (headset); .73–.86 across the paper and VR formats in Parkinson's disease (19 per format; about 17 days) & Parkinson's disease vs controls in time, errors and kinematics; age-group AUC close to paper; fatigue-related rise in interference cost in residents (abstract level) & none; none \\
Motion-capture CTT \parencite{ref60} & SUS and CSQ-VR measured & paper CTT: ρ .613 (A), .367 (B); n = 50, aged 19–24 & n/r & n/r & none; none \\
AR-TMT \parencite{ref61} & stage ratings of load, distraction and comfort; no usability or symptom scale & monitor TMT, always administered first: r .04 (A), .45 (B); n = 34 & n/r & n/r (distraction effects within participants) & none; none \\
VR-TMT for epilepsy \parencite{ref63} & five ad hoc items; one moderate symptom report in 8 & none (descriptive pilot) & n/r & n/r & none; none \\
Immersive TMT in cancer care \parencite{ref64} & 9 of 17 eligible completed & none & n/r & pre–post change only & none; none \\
VR TMT, technical report \parencite{ref69} & n/r & conventional TMT: significant for A and B (p < .05); coefficients not checked & n/r & n/r & none; none \\
TMT-VR, Varjo Aero PC \parencite{ref71,ref72,ref73} & SUS 79 of 100 \parencite{ref71}; reported mean 41.21, scoring requires clarification \parencite{ref72}; symptoms absent to mild & paper TMT (eye mode): r .36 (A), .28 (B); n = 53 & n/r & ADHD vs controls: no time difference; more wrong selections in ADHD & none; none \\
TMT-VR, English PC \parencite{ref74,ref75} & SUS 91 (younger), 87.4 (older); acceptability 4.85 (younger), 4.80 (older) of 6; CSQ-VR median 7 of 42 & paper TMT: r .51 (A), .51 (B), .36 (B−A); errors near zero; n = 91 & n/r & older vs younger, VR minus paper ΔAUC: Part B time .18, adjusted time .19, errors .22; Part A .04 (−.06 to .14) & none; none \\
TMT-VR, Greek standalone \parencite{ref76,ref78,ref79} & CSQ-VR ≤ 12 of 42 in 88.6\% of 202; 8.95 before and 7.83 after in 84 & paper TMT: Part B r .49 (comparison), .32 (MCI); Part A .38, .04; n = 84 & n/r & MCI vs comparison: Part B time g 1.02, errors g 1.08, Part A time g 0.05; highest AUCs B−A adjusted time .844, B−A time .816, Part B adjusted time .803; no format difference after correction; M-ACE .956 & none; no incremental value over the M-ACE shown \\
\bottomrule
\end{tabular}
\end{table*}

\subsection{Trail-Making Content and Training Outcomes}\label{sec:3.3}

The training question has four levels, and the sources answer them unevenly (Table 3). The first is whether XR training contains trail-like content and whether that content is acceptable. One headset game does. In the Darts game of the GAME2AWE platform, older adults burst numbered and lettered balloons in order by holding a central crosshair on each balloon until the arrow is released \parencite{ref67}. Its feasibility study reported usability and symptoms for the platform as a whole and no performance data for any game. By contrast, the trail-making mode of an AR music-making application is an embedded assessment of attention, not training content \parencite{ref66}.

The second level is whether programmes that contain such content change cognitive outcomes. A pilot randomised trial of the GAME2AWE platform found exploratory improvements on both parts of the paper TMT against usual care \parencite{ref68}. The platform combined Kinect, smart-floor and VR games, the combination of games varied by session and difficulty adapted automatically. The supplement that details the games was not accessible to this review, so exposure to the Darts game could not be verified. In a cancer-rehabilitation study, immersive TMT-A and TMT-B served as pre–post outcomes of VR cognitive rehabilitation delivered in the same system, in nine patients without a comparator \parencite{ref64}. The overlap between the trained tasks and the outcome task is unclear, so practice and regression to the mean cannot be separated from training. Beyond trail-like content, VR interventions for MCI gave small post-treatment effects in several domains \parencite{ref16}. In another meta-analysis of VR training in MCI, pooled effects favoured training on Part A and not on Part B \parencite{ref83}, and mixed-immersion trials gave pooled changes of a few seconds on the paper TMT \parencite{ref84}. The exergame evidence for executive function is small and heterogeneous \parencite{ref85}. These reviews overlap in their included trials and are not independent replications.

The third level is attribution. No identified trial isolates the efficacy of an immersive trail-making component. Immersion itself has not been isolated either. A network meta-analysis gave larger point estimates for non-immersive and semi-immersive than for immersive interventions against attention control, albeit by comparing different packages rather than by manipulating immersion \parencite{ref17}.

The fourth level is transfer to outcomes that were not trained. A gain on the paper TMT after practice on a trail-like game would be near transfer within the trail-making family. Broader benefit requires untrained measures of other cognitive functions and of everyday activity, with follow-up. None of the identified studies combines an active comparator, a defined dose of trail-making content, untrained outcomes at both distances and follow-up. Training software is also rarely shared \parencite{ref86}. Nonetheless, head-mounted exergames have been feasible and tolerated by older adults when sessions were short, supervised and preceded by acclimatisation \parencite{ref67,ref87,ref88}. The conditions for trials that can attribute a benefit to trail-making practice in XR are therefore in place.

\begin{table*}[!tp]
\caption{Trail-making content in XR training and the evidence available at each level of the training question. RCT, randomised controlled trial; SMD, standardised mean difference.}\label{tab:3}
\centering\scriptsize\setlength{\tabcolsep}{3pt}\renewcommand{\arraystretch}{1.12}
\begin{tabular}{L{0.2000\dimexpr\linewidth-18pt\relax}L{0.5000\dimexpr\linewidth-18pt\relax}L{0.3000\dimexpr\linewidth-18pt\relax}}
\toprule
\textbf{Level} & \textbf{Evidence identified} & \textbf{Boundary} \\
\midrule
Trail-like content and its acceptability & Darts: 12 numbered and lettered balloons burst in order with a head-gaze crosshair on Meta Quest 2; platform SUS 84 in a pre-pilot group of 4 experts and 8 older adults; mild symptoms in older end users \parencite{ref67} & one game in a 15-game platform; order rule and dwell time not specified; no game-level data \\
Programme outcomes & GAME2AWE pilot RCT vs usual care, 58 analysed, 16 weeks, cognitive outcomes exploratory; adjusted between-group differences: TMT-A −7.7 s (95\% CI −11.9 to −3.5), TMT-B −24.2 s (−35.6 to −12.8) \parencite{ref68}. Immersive TMT before and after VR rehabilitation in 9 patients \parencite{ref64}. VR in MCI, 18 RCTs, 722 adults: memory SMD .20 (.02 to .38), attention and processing speed .25 (.06 to .45), executive function .22 (.02 to .42), global cognition .09 (−.07 to .25) \parencite{ref16}. Part A change without Part B change \parencite{ref83}; pooled TMT changes of about 7–8 s \parencite{ref84}; exergames and executive function g = .05 (−.37 to .46), I² = 81\%, 11 studies \parencite{ref85} & Kinect, floor and VR games in a 5:3:2 usage ratio; Darts exposure not verifiable; no comparator and unclear task overlap in the rehabilitation study; reviews overlap in their trials \\
Attribution to an immersive trail-making component & none identified. Packages against attention control: immersive .29 (0 to .57), non-immersive .71 (.18 to 1.24), semi-immersive .74 (.44 to 1.04) \parencite{ref17} & different packages; not a randomised manipulation of immersion \\
Untrained transfer & paper TMT outcomes after a package that may include trail-like content \parencite{ref68} & near transfer within the trail-making family; broader outcomes and follow-up not combined with a comparator and a defined dose \\
\bottomrule
\end{tabular}
\end{table*}

\subsection{Comparative Appraisal and Remaining Gaps}\label{sec:3.4}

Table 4 condenses the common appraisal, and Supplementary Tables S7.1 and S7.2 give the matrices by criterion for every implementation. The same criteria show that the implementations differ more in what they have tested than in what they report. Most reports describe the display, task and selection method, but reporting remains incomplete for several implementations, most often for posture, error handling or build identity. None reports a check of its completion times or error counts against an external clock or observer. The samples are mostly convenience samples of young or healthy adults, and the clinical evidence comes from case–control designs.

Four gaps remain across the lineages, each bounded by the sources identified. First, reliability rests on one lineage tested on a fixed layout. Second, no implementation has normative data, a prospective validation in a referred population or a demonstrated increment over an existing screening instrument. Third, depth and error handling differ between lineages and have not been isolated as sources of score variance. No randomised-layout test has yet separated the search for a target from the movement to it. Only the VR-CTT has validated such a segmentation, against manually scored eye-tracking video, although its authors note that the two components overlap \parencite{ref47}. Fourth, code availability is uncommon. The AR-TMT paper links a public repository \parencite{ref61}, the VR-CTT offers data on request, and the source code of the 2026 TMT-VR is released with this review under the MIT licence (Data Availability Statement). A repository link does not by itself establish a licence, a deployable build or clinical readiness.

Applicability to older adults rests on two lineages. Among the implementations with validation data, only the VR-CTT and the TMT-VR have evaluated adults over 65, and only the TMT-VR has evaluated older adults with MCI. The AR-TMT, the motion-capture CTT and the epilepsy VR-TMT have been evaluated in young or middle-aged adults. Within the two lineages, the older participants were mostly healthy or cognitively screened volunteers, and no assessment study included people with dementia. The older TMT-VR participants were recruited through day centres for older adults \parencite{ref78,ref79}, and the Darts feasibility study took place at an elderly care centre \parencite{ref67}. The older-adult evidence therefore covers the healthier part of the population that these tests are meant to serve.

The fixed-layout tests cannot answer the reliability question for randomised forms, and the randomised-layout tests have not yet reported reliability. The TMT-VR case of Section 4 shows what access to source code and project records adds to that question.

\begin{table*}[!tp]
\caption{Compact appraisal of the XR trail-making lineages against the common criteria (the full matrices by criterion are Supplementary Tables S7.1 and S7.2). Implementations without evaluation data, or known only from an abstract, are appraised in Table S7.2 only.}\label{tab:4}
\centering\scriptsize\setlength{\tabcolsep}{3pt}\renewcommand{\arraystretch}{1.12}
\begin{tabular}{L{0.1700\dimexpr\linewidth-36pt\relax}L{0.1650\dimexpr\linewidth-36pt\relax}L{0.1650\dimexpr\linewidth-36pt\relax}L{0.1650\dimexpr\linewidth-36pt\relax}L{0.1700\dimexpr\linewidth-36pt\relax}L{0.1650\dimexpr\linewidth-36pt\relax}}
\toprule
\textbf{Lineage (reports)} & \textbf{Reporting completeness} & \textbf{Study method} & \textbf{Measurement evidence} & \textbf{Technical verification} & \textbf{Human factors and applicability} \\
\midrule
VR-CTT (9, one at abstract level) & device, task, layout principles and practice reported; no build identifier & healthy adults 18–90, Parkinson's disease and medical residents; short forms derived retrospectively & convergence; retest on a fixed layout; known groups in Parkinson's disease; no norms & reach segmentation checked against eye-tracking video; no timing check reported & standing with a controller; no usability or symptom questionnaire; 2 of 104 participants, both older adults, did not complete \\
TMT-VR (9) & configuration reported per study; build identifier in three reports; layouts not logged & ADHD and MCI comparisons; the MCI study has an asymmetric reference standard and an education imbalance & convergence; age-group discrimination; known groups in ADHD and MCI; no reliability; no norms & source examined for this review; placement defect found; timing not checked against an external clock & SUS, user experience, acceptability and CSQ-VR measured; seated and standing; adults to 90 years and MCI \\
AR-TMT (2) & device, geometry constraints and distraction conditions reported; public repository linked & 34 young adults; distraction varied within participants & convergence for Part B only; no reliability & not reported & stage ratings of comfort; no usability or symptom scale; young adults only \\
Motion-capture CTT (1) & device and tasks reported & 50 young adults & convergence (Spearman) & not reported & SUS and CSQ-VR in young adults \\
VR-TMT for epilepsy (1) & device, geometry and tutorial reported & descriptive pilot in 8 adults & none & not reported & ad hoc items; epilepsy intended, not yet studied \\
Darts game, GAME2AWE (2) & platform reported; game details in an inaccessible supplement & feasibility study; pilot trial of a multi-game package & diagnostic criteria not applicable; no game performance data & not reported & platform usability and symptoms in older adults at a care centre and in the community \\
Immersive TMT in cancer care (1) & implementation referred to earlier papers & uncontrolled pre–post study & pre–post change only & not reported & hospitalised adults aged 18–65; no symptom data \\
\bottomrule
\end{tabular}
\end{table*}

\section{The TMT-VR as a Measurement Case}\label{sec:4}

The TMT-VR is examined here because its source code, compiled builds and project records could be inspected for this review. The case shows what such access adds to the reading of a trail-making score and what it leaves unresolved.

\subsection{Configurations and Evidence Lineage}\label{sec:4.1}

The TMT-VR has nine reports on three configurations (Table 1). Three studies used a PC-tethered Varjo Aero to compare eye-tracking, head-gaze and controller ray-casting selection \parencite{ref71,ref73} and to examine the eye-tracking mode in adult ADHD \parencite{ref72}. The English configuration, tested standing on a VIVE Pro Eye, supports a usability report \parencite{ref74} and a report on convergence and age sensitivity \parencite{ref75}. The current version is a standalone application on a VIVE Focus Vision, administered seated in Greek, and it supports four reports \parencite{ref76,ref77,ref78,ref79}.

The current version selects by head-gaze: a reticle at the centre of view registers a target after a continuous 0.5-s dwell, and the dwell restarts if the reticle leaves the target. The dwell duration was set during the development of this version, after a pilot in 12 young and middle-aged adults. In that pilot 0.5 s performed better than longer durations on the System Usability Scale, which was the main criterion, and in oral feedback on the complexity of the task and on fatigue. The reports of the first configuration give the dwell as 2 s or 1.5 s. The 2024 and 2025 project files set the confirmation time to 0.5 s on every target handler, with the restart on exit. The published descriptions of the 2024–2025 studies are those of the papers, and the build each of those studies used was not archived with its data. The versions are therefore presented as distinct.

The configurations also differ in ways that change scores. The first two timed a part from its activation to the 25th selection, whereas the current version times from the first confirmation to the 25th, so that 24 links are timed. Posture changed from standing to seated, Part B uses Latin letters in English and Greek capitals in Greek, and the target field differs between configurations (Supplementary Table S4.1). In the current version, each part has an eight-target tutorial with prompts, an unrecorded 25-target practice run and the recorded test without prompts.

\subsection{Supported Findings and Measurement Limits}\label{sec:4.2}

Feasibility is the best-supported finding. In the reports of the two later configurations, all participants completed both parts, no session was stopped for symptoms, and symptoms after headset removal were low (Table 2). In the lifespan study, one adult in ten stood up once during a part to locate a target, with the clock running \parencite{ref78}. Eye tracking adds a condition. In the selection-technique comparison, some volunteers were excluded after initial calibration failure, and the analysed sample comprised young and middle-aged adults \parencite{ref73}. The calibration-failure rate in older adults therefore remains unknown. Usability was also measured in the three earlier reports \parencite{ref71,ref72,ref73} and in the English cohort. There, adults aged 64–81, most reporting little or no prior VR use, rated the task as usable and acceptable, with ratings that did not differ from those of younger adults after correction \parencite{ref74}.

The lifespan age findings come from the standalone cohorts. Completion time rose by 0.36 SD per decade (95\% CI 0.32 to 0.39) on a standardised Yeo–Johnson-transformed scale, with no quadratic term, and wrong-target errors rose by a factor of 1.24 per decade (1.17 to 1.33) \parencite{ref78}. The transformed scale does not imply constant increments in raw seconds, and the marginal R² of .54 belongs to age and part together. Neither lifespan sample showed an Age × Part interaction in time \parencite{ref76,ref78}. B−A time kept a smaller age association (β = .24, .11 to .36; Holm p = .003), whereas no B/A association was detected (β = .02, −.10 to .15). B−A is therefore not age-neutral, and the null for B/A does not show that B/A is. In the Fitts task, the proportional age slope was steeper for controller direct touch (6.9\% per decade, 5.1 to 8.7) than for ray-casting (3.8\%, 2.2 to 5.6) or head-gaze (3.3\%, 1.6 to 5.0) \parencite{ref76}.

The MCI evidence comes from one case–control study \parencite{ref79}. Part B scores separated adults with MCI from comparison adults, whereas head-gaze Part A time did not (Table 2). The Group × Part interaction survived correction for completion time (Holm p = .003) and error-adjusted time (p < .001), and not for error counts (Holm p = .070). The paper Part A separated the groups (g = 0.57, .14 to 1.00), albeit with an uncertain difference between formats (ΔAUC −.121, −.276 to .033; p = .125). The lower manual demand of head-gaze selection is one possible explanation among the differences between formats. Point estimates of discrimination favoured the TMT-VR's error and Part B-derived scores, although no difference between formats survived correction across 22 paired comparisons. The Mini-Addenbrooke's Cognitive Examination (M-ACE) discriminated better than any single score. Given an education difference of four years, an asymmetric reference standard and no reliability estimate, the TMT-VR remains a candidate trail-making format for MCI neuropsychological batteries pending prospective validation.

Convergence held for time scores and not for error counts. Within the English age groups, the Part B time correlation survived correction in both the younger (.43; Holm p = .024) and the older group (.48; Holm p = .010), and the Part A and B−A correlations did not \parencite{ref75}. Error counts did not converge. The paper examiner corrects each error as it occurs, whereas the application registers every completed dwell on a wrong target. In the MCI comparison, the observed maxima were 4 errors on paper and 22 in VR \parencite{ref79}. The differing recording rules may contribute to the lack of convergence. Ratios of group means between head-gaze and paper times were about 3.0 for Part A in the healthy groups and 2.6 in MCI, and 1.5–2.1 for Part B \parencite{ref75,ref79}. They are not individual ratios or conversion factors, and paper norms do not transfer to the TMT-VR.

Source inspection explains part of these patterns. Each of the 24 timed links includes a nominal 0.5-s confirmation, which adds a nominal 12 s to each part, and an interrupted dwell adds time that is not logged. A wrong-target error is a completed dwell on an unselected target other than the next target in the sequence. It is counted from the start of the part, so the counting window begins before the clock. A pause on a wrong target during search therefore completes a selection, and the error count can combine sequence errors with interaction-related selections. Error-adjusted time, the completion time multiplied by (1 + errors/24), depends on both.

The record has a larger gap. Every run generates a new layout, and no layout was stored. Path length varies between layouts with a coefficient of variation of about 10\%, so each score contains a layout component of unknown size (Supplementary Section S6). The examined placement routine also used the candidate field unevenly, favouring the nearest and farthest depth planes. It shuffles with a sort that uses a random comparison, which violates the sort's contract and occasionally leaves a part without targets. Neither layouts nor such failures were logged, so whether any past session met the defect cannot be established, although a part without targets would have been visible to the examiner. Layouts are independent of age by construction, so additive, age-independent layout effects add variance to age associations rather than bias. Because the layouts of the two parts are drawn independently, B−A carries the layout variance of both parts. By contrast, the nominal dwell contribution cancels from B−A and not from B/A.

\subsection{Implications for Reproducibility and Interaction Design}\label{sec:4.3}

A reproducible XR trail test must record each run's layout and seed, an event log of timestamped target entries, exits, interrupted dwells and completions, a head-pose log, and a session manifest that identifies the build, configuration and language. Scores can then be recomputed from events, the clock checked against an external clock, and the layout component of a score measured. For the TMT-VR, a seeded repair of the placement routine, these logging components and a bank of 4,000 candidate forms had been written and checked outside the Unity Editor. These components had not been integrated into the application or tested in the Unity Editor or on a headset as of 26 October 2026 (Supplementary Section S5). The repair is designed to reproduce every placement from a logged seed, including recovery from the sort failure. It did so for every tested seed in emulation, which remains to be shown on the deployed backend.

The candidate forms were matched on path length, summed angular separation and a Fitts-type index. Their coefficients of variation on these summaries are below 1\%, against 10\%, 13\% and 5\% among fresh layouts, although selection shifted other properties, such as vertical extent and the share of far targets (Supplementary Section S6). Geometric matching therefore reduces variation in the matched summaries, albeit without establishing equivalent difficulty for participants, alternate-form reliability or retest reliability. Fresh layouts and bank forms are two administration distributions and need distinct configuration identities. Recorded form histories, with verified exclusion of forms already seen, would reduce re-exposure to a specific path, although procedural, strategy and expectancy effects would remain. Such records would allow a retest and alternate-form study that estimates reliability and the variation due to persons, sessions and forms, and that evaluates differences between forms against prespecified equivalence margins.

Interaction design is the second implication. As currently exported, neither the paper TMT nor the head-gaze TMT-VR separates the time spent finding the next target from the time spent reaching and confirming it. A computer-administered TMT has separated search from movement, with age effects on both \parencite{ref32}, and eye and hand trajectories on a screen index separable processes \parencite{ref41,ref44}. The VR-CTT records have begun to do so in XR (Section 3.2), and kinematic markers carried discrimination in VR screening for MCI \parencite{ref89,ref90}. A VR adaptation of the Deary–Liewald reaction-time task \parencite{ref91} used the headset's eye tracker to report operational attention and motor intervals within choice responses, whose reliability and validity as separate components were not established \parencite{ref92}.

These precedents motivate a hypothesis rather than a design claim. Psychomotor performance is itself relevant to cognitive ageing, as age differences in simple and choice reaction times illustrate \parencite{ref91}. The paper TMT has a psychomotor component, because it combines visual search and cognitive control with manual movement \parencite{ref21,ref22,ref23}. Controller ray-casting adds a manual component absent from head-gaze selection, although it does not reproduce pencil tracing. Ray-casting is the candidate because the targets lie beyond reach and direct touch showed the steepest proportional age slope \parencite{ref76}. Continuous head and controller logs could therefore serve two aims. They could retain psychomotor performance that may be informative in its own right, and help examine how movement execution and cognitive control contribute to the score, especially in Part A.

The logs would relate movement paths, target visits and interrupted dwells to completion time. Where the final uninterrupted entry into a target is recorded, the interval before that entry can be distinguished from the confirmation interval that follows it. These intervals are operational. The first still combines search, decision and movement, and movement itself involves planning and online correction. Head orientation is not gaze, and neither interval measures pure motor speed. Validated segmentation, reliability estimates and a suitable motor baseline would be needed to investigate their contributions. A resulting motor index could first be examined alongside completion time and as a covariate, without assuming that adjustment leaves a pure cognitive score. Whether these records clarify age and clinical-group differences in Part A is the empirical question, and its answer would inform both assessment and the design of training tasks.

\section{Discussion and Implications}\label{sec:5}

The evidence reviewed here shows XR trail-making tests to be promising instruments, albeit with partial evidence for any clinical or training use. The TMT-VR case adds that the interaction technique, the confirmation and timing rules, the error definition and the layout record shape what such scores mean. For neuropsychologists, the requirements before clinical use are the familiar ones, applied to a new instrument \parencite{ref3}. Reliability must be estimated on the forms actually administered, which for randomised layouts means retest and alternate-form designs \parencite{ref50}. Scores need their own norms from prospective stratified samples. Diagnostic accuracy needs prospective validation in referred populations, with the reference standard applied to everyone and the increment over existing screening instruments tested. Error scores need definitions that separate sequence errors from interaction-related selections before they can support clinical decisions. Performance validity needs either a validated built-in indicator or a recommendation of concurrent validated measures \parencite{ref3}. Samples should also reach the older adults these tests are meant for, including adults over 80, people with MCI or dementia, care-home residents and people with sensory or motor impairment, with education sampled broadly.

For HCI researchers, the field and the case point to what should be logged, verified and reported. Every study should state a configuration identity: device, build, input and confirmation rule, timing boundary, posture and language. The layout of every run and an event log should be stored, and the error rule should state whether wrong inputs are registered, counted, corrected or ignored. Timing should be verified against an external clock, and symptoms should be measured with their time window stated, inside the headset as well as after removal where possible. Interaction choices change what is measured in older users. Head-gaze dwell avoids buttons, albeit at the cost of neck movement and of wrong selections completed by pauses during search. A controller ray adds manual coordination, direct touch showed the steepest relative age slope, and depth adds selection time. These choices are therefore part of the construct and should be tested with intended users rather than settled by convention.

The trail-making loop is also a candidate training mechanic. Each link requires a search under a rule, the inhibition of distractors, a movement and a confirmation, and Part B adds a switch between rules. These demands are cognitive and motor at once, which suits exergames for older adults but complicates attribution. One concept that follows is a seated game in which players release laser or water pistols at balloons or similar targets in sequence. Such a game would need the evidence that Section 3.3 found missing, and its design should begin with intended users. A seated exergame for persons living with dementia, for example, used no buttons and a recorded therapist voice as a guide \parencite{ref93}. That precedent supports button-free interaction with guidance; it does not establish a universal input rule. Trigger confirmation and drifting targets are therefore candidate choices to be tested. The available Fitts data cannot show that a trigger is faster than dwell for older adults, because every mode in them used dwell \parencite{ref76}. Moving targets carry known costs as well. In one co-designed exergame, approaching targets were replaced with static ones after older participants found their pacing difficult \parencite{ref81}. In another, adults around 80 caught fewer targets in a headset than in a projection-based room, although the headset gave greater presence \parencite{ref94}. The nearest precedent for the mechanic, a seated sword game with Stroop rules, was played by 12 older adults for 20 supervised sessions without a control arm \parencite{ref95}.

Dose needs descriptors that can be reported and replicated: the amount and range of movement, its intensity and the session duration, rather than duration and perceived exertion alone. The training game and the assessment should also remain separate configurations. A game may adapt spacing, tempo and difficulty, whereas the assessment keeps its rules and configuration identity, so that a change after training is not confounded with a change to the instrument. A comparator, untrained outcomes and follow-up are required. Gains on paper or VR trail tests after trail-like practice should be expected as near transfer, unless a broader outcome is justified in advance. Adaptive difficulty raises a further question. In the English assessment cohort, ratings of the experience were not associated with performance \parencite{ref74}. Such cross-sectional nulls justify neither including nor excluding frustration as an input to adaptation, which needs training-specific evidence. Colour is a design question too. Colour rules and feedback should be tested for discrimination and legibility with intended players, including older adults with cognitive impairment and people with colour-vision deficiencies, on the actual display. An immersive trail-making game is therefore a testable proposal, and trials that isolate its contribution will decide its value.

Overall, XR can make trail making more controlled, more richly recorded and more engaging for older adults. These advantages become evidence when the configuration is reported, the record is complete and each claim is tested in the population and design it concerns.

\section{Limitations}\label{sec:6}

This review applies the same criteria to every implementation and keeps lineages, configurations and studies apart. Its case rests on inspected source code and project records rather than on published descriptions alone. Its limitations indicate five directions for future work. First, one author screened and charted the records, without duplicate screening or logged queries, so implementations may have been missed; a registered scoping review with logged searches and duplicate screening should follow. Second, several sources were available only as abstracts or records, including several implementation reports, a trial supplement and a recent preprint attributed to the VR-CTT lineage. Obtaining them could change the inventory and the training appraisal. Third, the implementations differ in display, selection, layout, error handling and population, so the comparisons are qualitative; a common reporting set for XR trail tests would make them comparable. Fourth, the author developed the TMT-VR, so its appraisal is a self-appraisal under common criteria, and an independent appraisal of the software is needed. Fifth, the build used in some studies was not archived with their data, so the link between a study and an examined build rests partly on project records; future studies should archive the build with the data.

\section{Conclusions}\label{sec:7}

Trail making in XR has moved beyond feasibility, albeit not yet to clinical use. Convergent associations with paper or computerised trail tests were reported for four implementations, retest reliability for one lineage on a fixed layout and clinical group differences for two lineages. Normative data, prospective clinical validation, reliability of randomised forms and training benefit attributable to an immersive trail-making component were not identified. The TMT-VR case shows why these gaps matter: the interaction technique, the confirmation and timing rules, the error definition and unlogged random layouts all shape the meaning of a score. Further development for assessment therefore requires reported configurations, logged layouts and events, reliability estimated on the forms administered and prospective validation in the populations targeted. Training claims also require trials that isolate the trail-making component and test untrained outcomes. Overall, older adults, including adults with MCI, have tolerated supervised XR trail-making sessions well, which makes these next steps feasible.

\section*{Supplementary Materials}
The following supporting information accompanies this preprint: Section S1, search record and coverage of the evidence map; Section S2, contextual non-XR trail tests (Table S2.1) and the charting schema (Table S2.2); Section S3, XR inventory by features (Table S3.1), by evidence and openness (Table S3.2) and candidate and abstract-level records (Table S3.3); Section S4, TMT-VR configurations and provenance (Table S4.1) and the 26 exported scores (Table S4.2); Section S5, evidence status of the TMT-VR components (Table S5.1); Section S6, layout characterisation, placement defect and candidate form bank; Section S7, appraisal of every implementation against the common criteria (Tables S7.1 and S7.2) and source-specific appraisal notes (Table S7.3); File S1, the evidence map (152 records, 37 fields); File S2, the XR inventory as a spreadsheet.

\section*{Author Contributions}
Conceptualisation, methodology, software, validation, formal analysis, investigation, data curation, writing—original draft preparation, writing—review and editing, visualisation, project administration and funding acquisition, P.K. The author has read and agreed to the published version of the manuscript.

\section*{Funding}
This project has received funding from the European Union's Horizon Europe research and innovation programme under the Marie Skłodowska-Curie grant agreement No 101149390 (GERIATRIC). Views and opinions expressed are those of the author only and do not necessarily reflect those of the European Union or the European Research Executive Agency. Neither the European Union nor the granting authority can be held responsible for them.

\section*{Institutional Review Board Statement}
Not applicable. This review involved no new data collection; the TMT-VR studies synthesised here were approved by the ethics committees named in their reports.

\section*{Informed Consent Statement}
Not applicable.

\section*{Data Availability Statement}
The evidence map and the supplementary tables are provided with this article. The de-identified datasets of the TMT-VR studies are available in their Zenodo records: \url{https://doi.org/10.5281/zenodo.22767741} \parencite{ref76,ref77}, \url{https://doi.org/10.5281/zenodo.22770007} \parencite{ref78}, \url{https://doi.org/10.5281/zenodo.22825095} \parencite{ref79} and \url{https://doi.org/10.5281/zenodo.23270971} \parencite{ref74,ref75}. The source code of the 2026 TMT-VR project is available under the MIT licence at \url{https://github.com/PanosCortese/TMT-VR} as release 1.1.0, archived on Zenodo (\url{https://doi.org/10.5281/zenodo.23751063}). This release was prepared after the TMT-VR studies reviewed here, and its builds are not those used in them. It contains head-gaze and controller ray-casting versions in Greek and English, with standalone and PC builds attached to the GitHub release, and no validation of the English or PC versions is reported. The generated layout samples, the candidate form bank and the verification outputs described in Section 4.3 have not been released.

\section*{Acknowledgments}
During the preparation of this work the author used Anthropic Claude (Claude Fable 5.1 and Claude Opus 5.5 in the Claude app, and Claude Code; October 2026) and OpenAI Codex (GPT-6 Astra; October 2026). These tools assisted with organising the literature records and extracting and verifying their data, coding the layout emulation and the verification scripts, language refinement, and manuscript drafting and editing. The author reviewed, verified, and edited all outputs and takes full responsibility for the content of this work. Generative AI tools are not authors and received no CRediT roles.

\section*{Conflicts of Interest}
The author developed the TMT-VR and is an author of the studies of it that are synthesised in Section 4. The author is a Topic Advisor of \textit{Virtual Worlds} and a member of the RATE-XR Expert Group, whose reporting guideline is used in Section 2. The funder had no role in the design of the review, in the collection, analysis or interpretation of the literature, in the writing of the manuscript or in the decision to publish.

\section*{Abbreviations}
ADHD, attention-deficit/hyperactivity disorder; AR, augmented reality; AUC, area under the receiver operating characteristic curve; CI, confidence interval; CSQ-VR, Cybersickness in Virtual Reality Questionnaire; CTT, Color Trails Test; HCI, human–computer interaction; ICC, intraclass correlation coefficient; MCI, mild cognitive impairment; M-ACE, Mini-Addenbrooke's Cognitive Examination; MR, mixed reality; RCT, randomised controlled trial; SMD, standardised mean difference; SUS, System Usability Scale; TMT, Trail Making Test; TMT-VR, Trail Making Test in Virtual Reality; VR, virtual reality; VR-CTT, VR Color Trails Test; XR, extended reality.

\section*{Author identifiers}
ORCID iD: P.K., \url{https://orcid.org/0000-0002-2914-1064}.

\printbibliography[title={References}]
\end{document}